\pdfoutput=1
\documentclass[11pt]{article}

\usepackage[preprint]{acl}
\usepackage{times}
\usepackage{latexsym}
\usepackage[T1]{fontenc}
\usepackage[utf8]{inputenc}
\usepackage{microtype}
\usepackage{inconsolata}
\usepackage{graphicx}
\usepackage{amsmath}
\usepackage{amssymb}
\usepackage{booktabs}
\usepackage{array}
\usepackage{multirow}
\usepackage{enumitem}
\usepackage{xcolor}
\usepackage{placeins}

\title{Schema-First Retrieval: Embedding Catalogs \\ for Natural Language Analytics}

\author{Adarsh Agrawal \\
  \texttt{adagrawal@cs.stonybrook.edu} \\
  \And
  Shashank Indukuri \\
  \texttt{sinduku1@depaul.edu} \\}

\newcommand{\ours}{Schema-First Retrieval}
\newcommand{\figref}[1]{\hyperref[#1]{Figure~\ref*{#1}}}
\newcommand{\tabref}[1]{\hyperref[#1]{Table~\ref*{#1}}}
\newcommand{\eqnref}[1]{\hyperref[#1]{Equation~\ref*{#1}}}
\newcommand{\secref}[1]{\hyperref[#1]{Section~\ref*{#1}}}

\begin{document}
\maketitle

\begin{abstract}
Enterprise text-to-SQL systems often fail before SQL is generated: the model receives the wrong schema context. Modern warehouses contain thousands of tables, abbreviated columns, informal metrics, hidden join conventions, and permission boundaries that are not captured by raw table names. We introduce \ours, a retrieval layer that embeds catalog metadata rather than warehouse rows. The system indexes five typed catalog objects, tables, columns, metrics, relationships, and query history, using object-specific text templates. At query time, it combines parallel vector search, lineage expansion, cross-encoder reranking, workload memory, and deterministic access-control gates before SQL generation. On CRUSH4SQL (1,534 questions), \ours{} reaches 96.4\% table recall@20 and cross-encoder reranking adds +11.1 points at column recall@10; against an equally-templated BM25 baseline, semantic retrieval is +32.8 points at table recall@5. On SEDE (857 questions), query history raises table recall@5 from 52.1\% to 92.3\%. On BIRD (96 questions), schema-first context reduces SQL execution errors from 15.6\% to 6.2\%, a 2.5$\times$ reduction. These results show that catalog selection is a first-class retrieval problem for natural language analytics, not a prompt formatting detail.
\end{abstract}

\section{Introduction}
\label{sec:introduction}

Natural language analytics is usually framed as a generation problem: given a question and a database schema, produce SQL. That framing hides the step that decides whether generation can succeed. A model that sees the wrong tables, the wrong join keys, or the wrong business definition can produce fluent SQL that is syntactically valid and analytically useless. Recent enterprise benchmarks make this failure mode visible. Spider 2.0 evaluates models on realistic database workflows that require metadata search, dialect awareness, and project context rather than compact academic schemas \citep{lei2025spider2}. Large-scale schema-linking systems such as LinkAlign, AutoLink, RASL, SchemaGraphSQL, Solid-SQL, and context-aware bidirectional retrieval likewise treat schema selection as an independent problem \citep{wang2025linkalign,wang2026autolink,eben2025rasl,safdarian2025schemagraphsql,liu2025solid,nahid2025rethinking}.

This paper makes a stronger systems claim: the retrieval target should be the catalog. Enterprise users do not ask for \texttt{amt\_net\_rev\_lcl}; they ask for revenue. They do not know that churn lives in a lifecycle table, that a regional rollup uses a specific dimension, or that a dashboard metric contains a default exclusion. Those mappings already exist in modern catalogs as descriptions, metric definitions, lineage edges, ownership metadata, and historical query patterns. \ours{} turns that catalog into the retrieval substrate for text-to-SQL.

This paper makes the following contributions:
\begin{itemize}[leftmargin=*,topsep=2pt,itemsep=2pt]
\item \textbf{A five-type catalog object taxonomy.} We define separate embedding templates for tables, columns, metrics, relationships, and query history. These objects answer different retrieval questions: tables anchor the domain, columns ground predicates and projections, metrics bridge business language, relationships protect joins, and query history captures workload memory.
\item \textbf{A multi-signal retrieval pipeline.} We search all object types in parallel, expand candidates through lineage, rerank with a cross-encoder, and assemble the retrieved evidence into SQL-ready context for the generator.
\item \textbf{Deterministic governance outside the LLM.} We keep policy enforcement outside the language model by applying deterministic catalog filtering and SQL rewriting rather than asking the model to remember access rules.
\end{itemize}

The empirical pattern is consistent across three benchmarks. On CRUSH4SQL, reranking adds +11.1 points at column recall@10 and the full system reaches 96.4\% table recall@20; the bi-encoder beats an equally-templated BM25 baseline by +32.8 points at table R@5. On SEDE, where workload memory is available, query history raises table recall@5 by +40.2 points. On BIRD, retrieved catalog context reduces execution errors by 2.5$\times$ relative to full-schema prompting. The common lesson is simple: reliable natural language analytics needs a retrieval layer built for how warehouses are documented, queried, and governed.

\section{Related Work}
\label{sec:related}

Recent text-to-SQL research increasingly separates context selection from SQL generation. Spider 2.0 raises the benchmark standard by requiring models to operate over real workflows, large schemas, multiple dialects, and external metadata \citep{lei2025spider2}. LinkAlign decomposes large-scale text-to-SQL into database retrieval and schema item grounding \citep{wang2025linkalign}. AutoLink frames schema linking as iterative exploration over a database environment and a schema vector store \citep{wang2026autolink}. RASL decomposes metadata into retrievable semantic units for massive catalogs \citep{eben2025rasl}. Extractive schema linking, SchemaGraphSQL, Solid-SQL, and bidirectional schema retrieval further show that focused schema selection improves generation by removing distractor schema evidence \citep{glass2025extractive,safdarian2025schemagraphsql,liu2025solid,nahid2025rethinking}.

\ours{} is closest to this retrieval-centered line of work, but differs in the unit of retrieval. Most systems retrieve tables, columns, or generated schema snippets. We retrieve typed catalog objects, including business metrics, relationship objects, and query history. This makes the catalog itself the interface between analyst language and physical schema. We also distinguish ourselves from generation-centric pipelines such as DIN-SQL, DAIL-SQL, and MAC-SQL \citep{pourreza2023dinsql,gao2023dailsql,wang2024macsql}, which focus on decomposed in-context learning, self-correction, and multi-agent SQL refinement once schema is given; \ours{} addresses the upstream problem of which schema to give them. The design also follows the broader movement in retrieval-augmented generation toward adaptive retrieval, structured evidence, and graph- or domain-specific retrieval layers \citep{sharma2025rag,upadhyay2026trec,edge2024graphrag}. The generation backends we rely on are large instruction-tuned foundation models of the kind documented in recent technical reports \citep{amazon2025novapremier}, but our contribution is orthogonal to the choice of generator. For text-to-SQL, the domain structure is not a collection of passages. It is a governed catalog.

The governance side of \ours{} is motivated by work showing that access control cannot be treated as a soft instruction. Permissioned LLMs study access enforcement when models operate over siloed enterprise data \citep{jayaraman2025permissioned}. Role-conditioned refusal work extends Spider and BIRD with role-based policies and finds that explicit verification improves policy behavior \citep{klisura2025role}. Participant-aware access control argues for deterministic authorization in enterprise AI systems, especially when retrieval supplies private context \citep{bhatt2025participant}. We adopt that systems stance: retrieval and SQL rewriting are deterministic components, while the language model is only one step in the query path.

\section{Schema-First Retrieval}
\label{sec:method}

\begin{figure*}[t]
\centering
\includegraphics[width=0.98\textwidth]{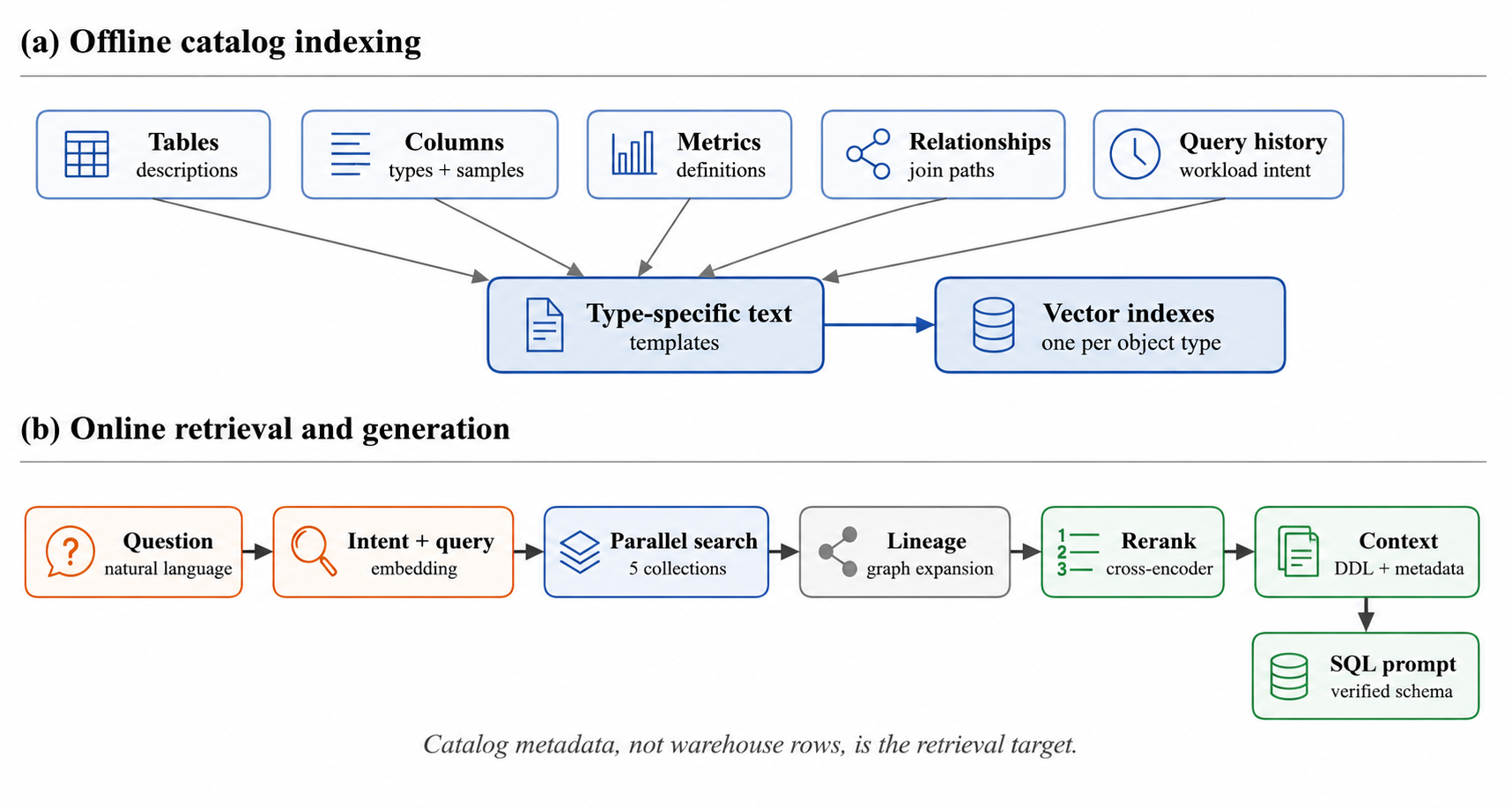}
\caption{\textbf{Schema-first retrieval pipeline.} Offline, five catalog object types are rendered with type-specific templates and embedded into separate vector indexes. Online, a question is classified and embedded, searched across catalog collections, expanded through the lineage graph, reranked with a cross-encoder, and assembled into compact schema context for SQL generation.}
\label{fig:pipeline}
\end{figure*}

\ours{} has an offline indexing phase and an online retrieval phase. Offline, each catalog object is rendered into a compact text representation and embedded into a type-specific vector collection. Online, the question is embedded once, searched against all catalog collections, expanded through lineage when joins are likely, reranked, and assembled into the context passed to SQL generation. \figref{fig:pipeline} shows the full path.

\subsection{Catalog Object Taxonomy}
\label{subsec:taxonomy}

We define five catalog object types. Tables carry names, schemas, descriptions, owners, and size cues; they are the primary anchors for retrieval. Columns add data types, parent table context, descriptions, and sample values; they ground projections, predicates, grouping keys, and filters. Metrics encode business definitions with formulas and source columns; they connect analyst vocabulary such as ``lifetime value'' or ``churn'' to concrete schema elements. Relationships encode foreign keys and documented joins; they reduce hallucinated join predicates by giving the generator verified paths. Query history stores prior natural language questions or query summaries with the tables and columns they accessed; it captures workload regularities that are invisible in static schema text.

The taxonomy is intentionally typed. A table description is useful for deciding where a question lives. A column sample is useful for deciding which predicate can express an entity. A metric definition is useful when business language does not appear in any table name. A historical query is useful when the same organization has asked a related question before. Rendering these objects with one generic template discards that structure, so each type receives its own embedding template. Appendix \secref{app:templates} gives the exact templates.

\subsection{Retrieval and Fusion}
\label{subsec:retrieval}

The first retrieval stage embeds the user question and searches the five catalog collections in parallel. We use Amazon Titan Embeddings V2 for 1,024-dimensional embeddings and ChromaDB for local vector storage. The initial candidate set is intentionally broad: the bi-encoder is responsible for coverage, not final ordering. When relationship evidence is needed, the lineage graph expands candidate tables through breadth-first join paths so that bridge tables and join keys can enter the context even if they are not directly named in the question.

The second stage reranks pooled candidates with Cohere Rerank V3.5. This cross-encoder stage is most important for columns, where names are short, duplicated across tables, and often abbreviated. The final retrieval score combines embedding similarity, reranker relevance, and query-history evidence:
\begin{equation}
\begin{aligned}
s(q, o) ={}& \alpha s_{\text{embed}}(q, o)
 + \beta s_{\text{rerank}}(q, o) \\
&+ \gamma s_{\text{history}}(q, o),
\end{aligned}
\label{eq:fusion}
\end{equation}
where $s_{\text{embed}}$ is normalized bi-encoder similarity, $s_{\text{rerank}}$ is cross-encoder relevance, and $s_{\text{history}}$ is derived from previous queries that accessed object $o$. The history signal is:
\begin{equation}
s_{\text{history}}(q, o) = \max_{h \in \mathcal{H}(o)} \cos(q, h) \exp(-\lambda \Delta t_h),
\label{eq:history}
\end{equation}
where $\mathcal{H}(o)$ is the set of historical queries that accessed object $o$, and $\Delta t_h$ is the age of historical query $h$. In \eqnref{eq:fusion}, a repeated workload pattern can promote the same schema while the system still falls back to semantic retrieval when no history exists.

\subsection{Generation and Access Control}
\label{subsec:generation_acl}

After retrieval, the context assembler emits DDL-style table schemas, metric definitions, and join paths. The SQL generator receives the user question, dialect instruction, and retrieved context, then produces SQL. Any instruction-tuned generator can fill this slot, including the multimodal reasoning, generation, and speech models that recent foundation-model families expose \citep{amazon2025nova,amazon2025nova2,amazon2025novasonic}; our pipeline treats the generator as a replaceable component. The generator is deliberately downstream of retrieval: it is not asked to discover the warehouse. It is asked to write SQL over a focused, catalog-validated subset.

Access control is handled as a deterministic system gate rather than a model instruction. Each catalog object can carry an access-control list specifying which users or groups may view it. Authorized context is defined as:
\begin{equation}
\text{Candidates}(u) = \{o \in \mathcal{C} \mid u \in \text{ACL}(o)\},
\label{eq:acl}
\end{equation}
where $\mathcal{C}$ is the catalog and $u$ is the requesting user. The generated SQL is then parsed and either rewritten with required row predicates or denied when it references a table outside the user's grants. \eqnref{eq:acl} captures the catalog-level filter, while the post-generation verifier protects the final SQL boundary.

\textbf{Implementation specifics.} Three components warrant brief mention. \textbf{(i) Intent-driven model routing.} A Haiku-based intent classifier tags each question with one of five categories (\texttt{single\_table}, \texttt{multi\_table}, \texttt{metric}, \texttt{temporal}, \texttt{exploratory}) and a binary \texttt{simple}/\texttt{complex} complexity label. Routing on a confidence-scored classifier follows the same data-driven, probability-learning principle used for adaptive decision-making in dynamic data-driven systems \citep{feng2024dddas}. The complexity label routes generation to Haiku for simple queries and Sonnet for complex ones, while the category biases retrieval; for example, \texttt{metric} queries up-weight the metric collection. \textbf{(ii) Token-budgeted context assembly.} Within a 5{,}000-token budget, the assembler allocates fixed fractions across object types (30\% tables, 35\% columns, 20\% query examples, 10\% metric definitions, 5\% join paths) and fills each bucket greedily in re-ranking order. The 35\% share for columns reflects the empirical finding that column-level retrieval is the dominant residual error mode (\secref{subsec:complexity}). \textbf{(iii) AST-based access control.} Row-level predicates are parsed via \texttt{sqlglot.condition()} into AST nodes and injected as \texttt{AND}-conjuncts on every \texttt{SELECT} that references a protected table, with CTE aliases excluded from the table-resolution pass. This avoids the parser-confusing failures of string-concatenation rewriters on queries with subqueries, joins, or shadowed alias names.

\begin{center}
\colorbox{black!6}{\parbox{0.93\columnwidth}{\small \textbf{Principle.}\;The retrieval target is the catalog. Tables, columns, metrics, relationships, and query history are typed objects that answer different retrieval questions; the language model writes SQL over a focused, catalog-validated subset, and access control is enforced deterministically outside the model.}}
\end{center}

\begin{figure}[t]
\centering
\colorbox{black!6}{\parbox{0.95\columnwidth}{\footnotesize
\textbf{Worked example.} BIRD question on the \texttt{california\_schools} database (difficulty: simple).

\smallskip
\textbf{Question.} \emph{``How many schools in Amador which the Low Grade is 9 and the High Grade is 12?''}

\smallskip
\textbf{Retrieved catalog (top 4 of 50).}
\begin{itemize}[leftmargin=1.2em,topsep=2pt,itemsep=1pt]
\item \texttt{Column: frpm."County Name"} \;|\; samples: \emph{Alameda, Amador, Butte} \;|\; type: \texttt{TEXT}
\item \texttt{Column: frpm."Low Grade"} \;|\; samples: \emph{K, 1, 9, P} \;|\; type: \texttt{TEXT}
\item \texttt{Column: frpm."High Grade"} \;|\; samples: \emph{12, 8, 6, Adult} \;|\; type: \texttt{TEXT}
\item \texttt{Table: frpm} \;|\; \emph{Free and Reduced Price Meal eligibility data} \;|\; 9{,}986 rows
\end{itemize}

\smallskip
\textbf{Full-schema SQL (failed).}\\
\texttt{SELECT COUNT(*) FROM frpm}\\
\texttt{~WHERE County Name = 'Amador'}\\
\texttt{~~~AND Low Grade = '9'}\\
\texttt{~~~AND High Grade = '12'}\\
\textcolor{red!70!black}{\textbf{Error:}} \texttt{near "Name": syntax error}

\smallskip
\textbf{Schema-First SQL (succeeded).}\\
\texttt{SELECT COUNT(*) FROM frpm}\\
\texttt{~WHERE "County Name" = 'Amador'}\\
\texttt{~~~AND "Low Grade" = '9'}\\
\texttt{~~~AND "High Grade" = '12'}

\smallskip
\textbf{Why retrieval helped.} The column template embeds each multi-word column name inside double quotes (\texttt{"County Name"}) so the rendered context already shows the safe quoting form. The generator copies this surface form, sidestepping the most common BIRD failure mode without any prompt instruction about quoting.
}}
\caption{\textbf{End-to-end worked example.} The retrieved column objects render with proper double-quoting. The schema-first generator copies this surface form and avoids the unquoted-identifier syntax error that breaks the full-schema baseline.}
\label{fig:worked_example}
\end{figure}

\section{Experimental Setup}
\label{sec:setup}

We evaluate three surfaces of the pipeline. CRUSH4SQL provides 1,534 natural language questions with gold table and column annotations, so it isolates schema retrieval quality \citep{luo2023crush4sql}. SEDE provides 857 Stack Exchange Data Explorer questions and naturally occurring query history, so it tests whether workload memory improves retrieval \citep{hazoom2021sede}; we use the dataset's predefined 10,309-query \emph{train} split as the history corpus and the 857-question \emph{test} split as evaluation, giving a strict train/test separation between history embeddings and queries scored. BIRD provides realistic database schemas and text-SQL pairs; we use a 96-question subset stratified across difficulty bands (58 simple, 32 moderate, 6 challenging) for end-to-end SQL generation \citep{li2023bird}.

For retrieval we report Recall@$K$, Mean Reciprocal Rank, NDCG@10, and MAP@10. For SQL generation we report execution accuracy, result accuracy, execution error rate, and a paired exact McNemar's test on the per-question pass/fail outcomes. The embedding model is Amazon Titan Embeddings V2, the reranker is Cohere Rerank V3.5, and the SQL generation experiments use the same generation model and prompt template in both compared conditions. Appendix \secref{app:implementation} lists implementation parameters and the complete hyperparameter table.

\section{Results}
\label{sec:results}

\subsection{Retrieval Quality}
\label{subsec:retrieval_results}

\begin{table*}[t]
\centering
\small
\begin{tabular}{llcccc}
\toprule
\textbf{Benchmark} & \textbf{Condition} & \textbf{Table R@5} & \textbf{Table R@20} & \textbf{Column R@10} & \textbf{Extra signal} \\
\midrule
CRUSH4SQL & BM25 (sparse) & 50.0\% & 63.4\% & 28.7\% & MRR 0.438 \\
CRUSH4SQL & Bi-encoder, no reranker & 82.8\% & 94.3\% & 44.4\% & MRR 0.624 \\
CRUSH4SQL & Bi-encoder + reranker & 89.1\% & 96.4\% & 55.5\% & MRR 0.761 \\
\midrule
SEDE & No history & 52.1\% & 83.5\% & 7.7\% & Single-table R@5 71.0\% \\
SEDE & + History & 92.3\% & 95.0\% & 7.7\% & Single-table R@5 98.4\% \\
\bottomrule
\end{tabular}
\caption{\textbf{Main retrieval effects.} BM25 over the same rendered catalog templates is included as an external sparse-retrieval baseline. On CRUSH4SQL, semantic retrieval beats BM25 by +32.8 points at Table R@5; cross-encoder reranking adds another +6.3 points at Table R@5 and +11.1 at Column R@10. On SEDE, query history supplies the strongest table-level signal, raising table recall@5 by +40.2 points while leaving the sparse SEDE column annotations unchanged. Full $K$ sweeps are in Appendix \tabref{tab:crush_full} and Appendix \tabref{tab:sede_full}.}
\label{tab:retrieval_main}
\end{table*}

\tabref{tab:retrieval_main} shows three complementary retrieval effects. \textbf{Sparse retrieval is a weak baseline.} BM25 over the same rendered catalog templates reaches only 50.0\% Table R@5 and 28.7\% Column R@10, $-$32.8 and $-$15.7 points relative to the bi-encoder alone. The gap quantifies the contribution of semantic similarity beyond lexical overlap and confirms that the gains do not come purely from the templates: the same text indexed with a sparse retriever is substantially worse. \textbf{Reranking helps columns most.} On CRUSH4SQL, the cross-encoder reranker improves every reported metric. The largest absolute gains appear at the column level: +10.5 points at column recall@5 and +11.1 points at column recall@10. This is the expected place for a cross-encoder to matter. Tables usually carry descriptions and domain names, but columns are short, duplicated across tables, and frequently abbreviated. Reranking turns the bi-encoder candidate pool into a usable ordering for SQL context.

\textbf{History moves the right table to the front.} SEDE shows a different effect. Adding query history raises table recall@5 from 52.1\% to 92.3\%, a +40.2-point gain. The result is not simply that history finds more tables eventually. At recall@20, the gap narrows to 95.0\% versus 83.5\% (+11.5 points), and at recall@50 to 96.8\% versus 91.4\% (+5.4 points). History therefore contributes ranking precision, not coverage: it achieves at $K{=}5$ what the bi-encoder alone requires $K{=}20$ to reach. The sparse SEDE column annotations (only 7.7\% column R@10 in both conditions) do not expose the same benefit at the column level, so we report this as a table-selection effect rather than a general column claim; Appendix \secref{app:full_results} gives the full sweep.

\begin{table}[t]
\centering
\small
\begin{tabular}{lcc}
\toprule
\textbf{Removed type} & \textbf{Table R@5} & \textbf{Col R@10} \\
\midrule
None (full) & 89.1\% & 55.5\% \\
$-$ Query history & 84.2\% & 52.1\% \\
$-$ Metrics & 85.0\% & 53.8\% \\
$-$ Relationships & 87.4\% & 52.3\% \\
$-$ Column objects & 88.3\% & 52.7\% \\
$-$ Table descriptions & 87.2\% & 54.1\% \\
\bottomrule
\end{tabular}
\caption{\textbf{Object type ablation on CRUSH4SQL.} Each row removes one of the five catalog object types. No type is redundant; relationship removal hurts column recall most because it removes join-key evidence.}
\label{tab:type_ablation_main}
\end{table}

\textbf{Every catalog object type carries non-redundant signal.} Removing query history degrades table R@5 by 4.9 points; removing metrics costs 4.1 points; removing relationships costs 3.2 points at column R@10. Even removing column objects, the type with the smallest contribution, still costs 0.8 table R@5 and 2.8 column R@10 points. The full range across the five removals is 0.8 to 4.9 points (table R@5) and 1.4 to 3.4 points (column R@10), so no single type dominates and none is redundant. The ablation validates the typed taxonomy as a design choice rather than a stylistic one.

\subsection{End-to-End SQL Generation}
\label{subsec:generation_results}

\begin{table}[t]
\centering
\footnotesize
\begin{tabular}{lccc}
\toprule
\textbf{Condition} & \textbf{Exec} & \textbf{Result} & \textbf{Errors} \\
\midrule
Full schema & 84.4\% & 34.4\% & 15.6\% \\
\ours{} & 93.8\% & 36.5\% & 6.2\% \\
\bottomrule
\end{tabular}
\caption{\textbf{End-to-end SQL generation on BIRD} (96 questions). Both conditions use the same generation model and prompt template. The only change is whether the model receives the full database schema or schema-first retrieved context. \ours{} reduces execution errors by 2.5$\times$.}
\label{tab:generation}
\end{table}

\textbf{Errors drop by 2.5$\times$.} \tabref{tab:generation} tests the retrieval layer inside SQL generation. \ours{} reduces execution errors from 15.6\% to 6.2\% (15 to 6 failures out of 96), a 2.5$\times$ improvement. The reduction is statistically significant under a paired exact McNemar's test on the per-question outcomes ($b{=}12$, $c{=}3$; two-sided $p{=}0.035$), with a 95\% CI of $[1.7, 17.1]$ percentage points on the error-rate reduction. Of the 9 eliminated failures, 8 are at the lexical-grounding layer (identifier quoting and wrong-table column references) as detailed in \secref{subsec:errors}. The result accuracy gain is smaller, +2.1 points (34.4\% to 36.5\%) and is not statistically significant (paired McNemar's $p{=}0.69$), because result accuracy also depends on aggregation, filtering, ordering, and other SQL reasoning choices after the correct schema is present. The execution result isolates the part of the error surface that schema selection can directly repair.

\begin{figure*}[t]
\centering
\includegraphics[width=0.95\textwidth]{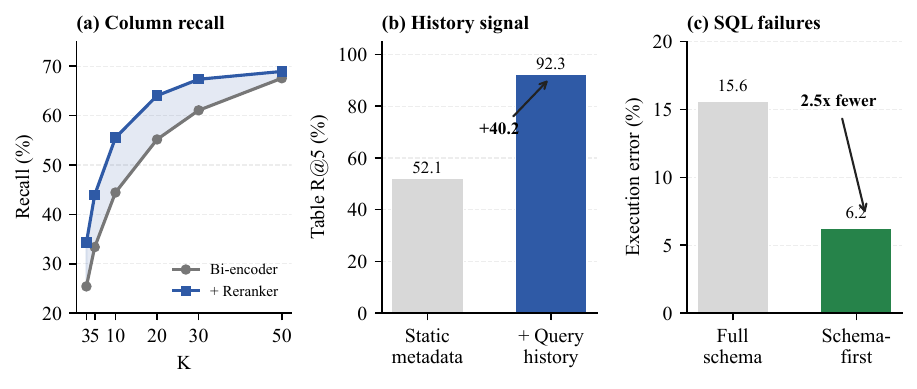}
\caption{\textbf{Main empirical effects across benchmarks.} Reranking improves CRUSH4SQL column recall across low-$K$ operating points, query history sharply improves SEDE table recall at R@5, and schema-first context reduces BIRD SQL execution failures from 15.6\% to 6.2\%.}
\label{fig:dashboard}
\end{figure*}

\figref{fig:dashboard} summarizes the same pattern visually. The important point is not that one signal dominates every benchmark. The important point is that the signals are complementary. Embedding provides broad semantic coverage. Reranking improves fine-grained candidate order. History supplies local workload memory. Relationship objects supply join evidence. The system works because the catalog is not treated as one undifferentiated text blob.

\subsection{Complexity and Failure Modes}
\label{subsec:complexity}

\begin{table}[t]
\centering
\small
\begin{tabular}{lcc}
\toprule
\textbf{Gold tables} & \textbf{Table R@5} & \textbf{Column R@10} \\
\midrule
1 table & 96.9\% & 73.4\% \\
2 tables & 88.5\% & 51.3\% \\
3 to 4 tables & 79.7\% & 45.0\% \\
\bottomrule
\end{tabular}
\caption{\textbf{Retrieval quality by query complexity} on CRUSH4SQL (with reranker). Column recall degrades more sharply than table recall as the number of required tables increases.}
\label{tab:complexity_main}
\end{table}

\textbf{Single-table queries are nearly solved.} \tabref{tab:complexity_main} sharpens the systems lesson. Single-table CRUSH4SQL queries are close to solved at the table level, with 96.9\% table recall@5 and 73.4\% column recall@10. With 3 to 4 gold tables, table recall@5 remains 79.7\% (a graceful $-$17.2-point decline) but column recall@10 falls to 45.0\% (a sharper $-$28.4-point decline). The fine-grained task is no longer finding the broad topic of the question. It is selecting the right projections, predicates, aggregations, and join keys across several related tables. This is why relationship objects and lineage expansion are part of the retrieval path rather than a cosmetic addition.

\textbf{Moderate queries gain most.} The BIRD difficulty analysis in Appendix \tabref{tab:bird_difficulty} shows that the largest execution-accuracy gain appears on moderate queries, from 75.0\% to 96.9\%. Simple queries often contain enough surface evidence for either condition. The challenging split is small in this subset, so we use it for transparency rather than as a separate claim.

\subsection{Error Analysis}
\label{subsec:errors}

\begin{table}[t]
\centering
\small
\begin{tabular}{lccc}
\toprule
\textbf{Failure type} & \textbf{Full} & \textbf{SFR} & \textbf{$\Delta$} \\
\midrule
Identifier quoting & 6 & 0 & $-$6 \\
Wrong-table column ref. & 2 & 0 & $-$2 \\
Column does not exist & 1 & 1 & 0 \\
Type / NULL comparison & 2 & 1 & $-$1 \\
\texttt{DISTINCT} aggregate arity & 1 & 1 & 0 \\
Other parser errors & 3 & 3 & 0 \\
\midrule
\textbf{Total} & \textbf{15} & \textbf{6} & \textbf{$-$9} \\
\bottomrule
\end{tabular}
\caption{\textbf{Error taxonomy on BIRD} (96 questions). Schema-first retrieval eliminates 8 of 9 fixed errors at the lexical-grounding layer (identifier quoting and wrong-table references); the remaining six failures are schema-orthogonal.}
\label{tab:error_taxonomy}
\end{table}

\textbf{Schema retrieval eliminates lexical-grounding errors.} \tabref{tab:error_taxonomy} groups the 21 execution failures across both conditions by root cause. Eight of the nine errors fixed by schema-first retrieval are lexical: six are identifier-quoting failures (e.g., the model emits \texttt{County Name} unquoted, breaking the SQLite parser), and two are wrong-table column references that retrieval disambiguates by surfacing the right table-column pairing in context. The retrieved column objects render names with the safe quoting form, which the generator copies. We acknowledge that a prompt-engineering fix (e.g., adding \emph{``quote identifiers containing spaces''} to the full-schema baseline) would address the six identifier-quoting failures and narrow the gap to roughly 1.4$\times$. The contribution should therefore be read as: \emph{templated catalog rendering propagates safe surface forms through retrieval, replacing per-prompt instruction with a structural property of the retrieved context}. \textbf{Schema retrieval cannot fix data-semantics errors.} The remaining six failures under schema-first are NULL-vs-string comparisons, aggregate-arity violations, and parser edge cases that no amount of better schema context can repair. This is the failure surface where retrieval is no longer the bottleneck.

\subsection{Efficiency}
\label{subsec:efficiency}

\begin{table}[t]
\centering
\small
\begin{tabular}{lr}
\toprule
\textbf{Phase} & \textbf{Latency} \\
\midrule
Intent classification (Haiku) & \;\;500\,ms \\
Query embedding (Titan V2)$^\dagger$ & \;\;200\,ms \\
Vector search (5 parallel) & \;\;\;\;\;5\,ms \\
Cross-encoder rerank (Cohere V3.5) & \;\;400\,ms \\
\midrule
Retrieval end-to-end (mean) & 1{,}280\,ms \\
SQL generation (downstream) & 1{,}500\,ms \\
\bottomrule
\end{tabular}
\caption{\textbf{Per-phase latency on CRUSH4SQL} (mean over 1{,}534 questions). $^\dagger$Embedding runs concurrently with intent classification, so retrieval end-to-end ($\approx$1.28\,s/query) is dominated by the slowest concurrent leg plus reranking. Sustained throughput is 0.93 queries/second.}
\label{tab:latency}
\end{table}

\textbf{The pipeline is interactive.} \tabref{tab:latency} reports per-phase latency: retrieval completes in roughly 1.28\,s/query at sustained 0.93 q/s on a single host with no caching, and SQL generation adds another ${\sim}1.5$\,s downstream. The dominant Bedrock calls are sequential, so optimizations such as warm caches for repeat queries and reranker batching directly translate into lower latency. \textbf{Total experimental cost was under \$3 across 5{,}316 questions} (CRUSH4SQL, SEDE, and BIRD combined), or about \$0.006 per question end-to-end at our model mix; SQL generation is the dominant cost item, with reranking and intent classification two orders of magnitude cheaper. \textbf{BIRD does not show token savings.} Average context tokens are 445 (full schema) versus 459 (schema-first) on BIRD because BIRD databases contain only 3 to 13 tables; full-schema prompts already fit in the budget. The schema-first win on BIRD is therefore quality (error reduction), not prompt-length reduction. Token savings scale with schema size and are visible at enterprise scale (thousands of tables, see \secref{sec:discussion}).

\section{Discussion}
\label{sec:discussion}

\textbf{The system boundary changes.} \ours{} changes the system boundary for natural language analytics. The language model is no longer responsible for discovering the warehouse, inferring undocumented joins, and remembering policy while writing SQL. Retrieval supplies a focused, provenance-carrying catalog slice. Generation writes over that slice. Verification and access control check the final boundary. This decomposition is the reason the same architecture can improve both retrieval metrics and execution behavior. The same separation of concerns appears in tool-augmented agent systems, where reliability is improved by isolating and self-healing failing components rather than relying on a single monolithic model \citep{babu2026selfhealing}; schema-first retrieval can be viewed as the catalog-grounding tool in such a pipeline.

The comparison to recent schema-linking work is instructive. LinkAlign and AutoLink focus on discovering relevant schema in large multi-database settings \citep{wang2025linkalign,wang2026autolink}. RASL and extractive schema linking focus on decomposing and retrieving schema components \citep{eben2025rasl,glass2025extractive}. SchemaGraphSQL adds explicit graph pathfinding for joins \citep{safdarian2025schemagraphsql}. \ours{} is compatible with these directions, but changes the representation being searched. The system retrieves business and governance objects from the catalog, not only physical tables and columns. That distinction matters most when the user's language is business language and the warehouse language is operational shorthand.

The governance result is architectural rather than rhetorical. The system never relies on a prompt sentence such as ``follow the access policy.'' Catalog filtering and SQL rewriting are explicit program steps. This design follows the direction of recent enterprise-AI access-control work: policy must be enforced by the surrounding system, then audited at the SQL boundary \citep{jayaraman2025permissioned,klisura2025role,bhatt2025participant}.

\section{Conclusion}
\label{sec:conclusion}

\textbf{The catalog is the semantic object that should be embedded.} We presented \ours, a catalog-centered retrieval layer for natural language analytics. Across CRUSH4SQL, SEDE, and BIRD, the same pattern holds: catalog-aware retrieval improves schema recall, query history supplies powerful workload memory, and focused schema context reduces SQL execution failures. Together, the five-type taxonomy and multi-signal pipeline move text-to-SQL from raw schema prompting toward governed catalog retrieval.

\section*{Limitations}

\textbf{Benchmark scope.} We evaluate public benchmarks that proxy enterprise natural language analytics rather than private production warehouses. CRUSH4SQL and SEDE are useful for isolating retrieval behavior, and BIRD is useful for end-to-end SQL generation, but real deployments also depend on catalog completeness, organization-specific metric definitions, and warehouse governance practices. \textbf{BIRD evaluation size.} The BIRD generation study uses 96 questions stratified across difficulty bands; while the 9.4-point execution-accuracy gain is significant under a paired McNemar's test ($p{=}0.035$), the challenging slice ($N{=}6$) is too small for an independent claim, and the result-accuracy gain ($+2.1$ points) is not statistically significant. Scaling to the full 1,534-question BIRD dev set is the natural next step. \textbf{Comparison to prior schema-linking systems.} We compare against within-system ablations and an external BM25 baseline rather than running LinkAlign, AutoLink, or RASL head-to-head on the same benchmarks. Direct comparison would require reproducing those systems on CRUSH4SQL, SEDE, and BIRD with consistent retrieval budgets; we leave this to future work and note that our contribution is the typed-catalog representation, which is compatible with and could be combined with those retrievers. \textbf{Catalog completeness.} Our benchmarks have curated descriptions, sample values, and (for SEDE) workload history; production catalogs are typically sparser. We probe noise robustness directly in Appendix \secref{app:noise} by dropping 25\%, 50\%, and 75\% of column and table descriptions; on CRUSH4SQL the system is essentially noise-immune because table and column NAMES carry most of the discriminative signal. We expect higher noise sensitivity on deployments with rich descriptions; this is a deployment-time consideration. \textbf{Workload memory dependence.} Query history is valuable when prior workloads are available and authorized for use; deployments without retained query logs fall back to static catalog retrieval. \textbf{Access control.} The deterministic filtering and SQL rewriting verifies the system design, but it is not a substitute for a full security audit of an enterprise deployment.

\section*{Ethical Considerations}

The system is designed to reduce hallucinated SQL and to enforce catalog permissions outside the language model. Its main risks come from metadata handling. Catalog descriptions, metric definitions, and query history can reveal sensitive business information even when warehouse rows are not exposed. Deployments should apply the same authorization, audit, retention, and minimization policies to catalog metadata and query logs that they apply to database access. Beyond access control, the language models that consume the retrieved context inherit the safety and bias considerations documented for general-purpose models; standardized safety benchmarks \citep{vidgen2024mlcommons} and bias-mitigation techniques \citep{agrawal2022debiasgan} are complementary safeguards that a production deployment should adopt alongside the catalog-level governance studied here. The experiments use public benchmark artifacts and do not involve human-subject data collection.

\bibliography{references}

\begin{thebibliography}{28}
\providecommand{\natexlab}[1]{#1}

\bibitem[{{Amazon Artificial General
  Intelligence}(2025{\natexlab{a}})}]{amazon2025nova2}
{Amazon Artificial General Intelligence}. 2025{\natexlab{a}}.
\newblock {Amazon Nova 2: Multimodal Reasoning and Generation Models}.
\newblock Technical report, Amazon.

\bibitem[{{Amazon Artificial General
  Intelligence}(2025{\natexlab{b}})}]{amazon2025novapremier}
{Amazon Artificial General Intelligence}. 2025{\natexlab{b}}.
\newblock \href
  {https://www.amazon.science/publications/amazon-nova-premier-technical-report-and-model-card}
  {{Amazon Nova Premier: Technical Report and Model Card}}.
\newblock Technical report, Amazon.

\bibitem[{{Amazon Artificial General
  Intelligence}(2025{\natexlab{c}})}]{amazon2025novasonic}
{Amazon Artificial General Intelligence}. 2025{\natexlab{c}}.
\newblock {Amazon Nova Sonic: Technical Report and Model Card}.
\newblock Technical report, Amazon.

\bibitem[{{Amazon Artificial General
  Intelligence}(2025{\natexlab{d}})}]{amazon2025nova}
{Amazon Artificial General Intelligence}. 2025{\natexlab{d}}.
\newblock \href {https://arxiv.org/abs/2506.12103} {{The Amazon Nova Family of
  Models: Technical Report and Model Card}}.
\newblock \emph{arXiv preprint arXiv:2506.12103}.

\bibitem[{Babu and Agrawal(2026)}]{babu2026selfhealing}
Rahul~Suresh Babu and Adarsh Agrawal. 2026.
\newblock \href {https://arxiv.org/abs/2606.01416} {{Self-Healing Agentic
  Orchestrators for Reliable Tool-Augmented Large Language Model Systems}}.
\newblock \emph{Preprint}, arXiv:2606.01416.

\bibitem[{Bhatt et~al.(2025)Bhatt, Rajore, Aggarwal, Ananthanarayanan, Chandra,
  Chandran, Choudhury, Gupta, Kiciman, Pandey, Setty, Sharma, and
  Zhao}]{bhatt2025participant}
Shashank~Shreedhar Bhatt, Tanmay Rajore, Khushboo Aggarwal, Ganesh
  Ananthanarayanan, Ranveer Chandra, Nishanth Chandran, Suyash Choudhury, Divya
  Gupta, Emre Kiciman, Sumit~Kumar Pandey, Srinath Setty, Rahul Sharma, and
  Teijia Zhao. 2025.
\newblock \href {https://arxiv.org/abs/2509.14608} {{Enterprise AI Must Enforce
  Participant-Aware Access Control}}.
\newblock \emph{Preprint}, arXiv:2509.14608.

\bibitem[{Eben et~al.(2025)Eben, Ahmad, and Lau}]{eben2025rasl}
Jeffrey Eben, Aitzaz Ahmad, and Stephen Lau. 2025.
\newblock \href {https://arxiv.org/abs/2507.23104} {{RASL: Retrieval Augmented
  Schema Linking for Massive Database Text-to-SQL}}.
\newblock \emph{Preprint}, arXiv:2507.23104.

\bibitem[{Edge et~al.(2024)Edge, Trinh, Cheng, Bradley, Chao, Mody, Truitt, and
  Larson}]{edge2024graphrag}
Darren Edge, Ha~Trinh, Newman Cheng, Joshua Bradley, Alex Chao, Apurva Mody,
  Steven Truitt, and Jonathan Larson. 2024.
\newblock \href {https://arxiv.org/abs/2404.16130} {{From Local to Global: A
  Graph RAG Approach to Query-Focused Summarization}}.
\newblock \emph{arXiv preprint arXiv:2404.16130}.

\bibitem[{Feng et~al.(2024)Feng, Agrawal, Ling, Blasch, Adiles-Cruz, Schrader,
  and Wei}]{feng2024dddas}
Wei Feng, Adarsh Agrawal, Haibin Ling, Erik Blasch, Edmund Adiles-Cruz,
  Philip~T. Schrader, and Jia Wei. 2024.
\newblock {DDDAS Probability Learning for Natural Disaster Change Detection}.
\newblock In \emph{International Conference on Dynamic Data Driven Applications
  Systems}, pages 90--99.

\bibitem[{Gao et~al.(2024)Gao, Wang, Li, Sun, Qian, Ding, and
  Zhou}]{gao2023dailsql}
Dawei Gao, Haibin Wang, Yaliang Li, Xiuyu Sun, Yichen Qian, Bolin Ding, and
  Jingren Zhou. 2024.
\newblock \href {https://arxiv.org/abs/2308.15363} {{Text-to-SQL Empowered by
  Large Language Models: A Benchmark Evaluation}}.
\newblock \emph{Proceedings of the VLDB Endowment}.

\bibitem[{Glass et~al.(2025)Glass, Eyceoz, Subramanian, Rossiello, Vu, and
  Gliozzo}]{glass2025extractive}
Michael Glass, Mustafa Eyceoz, Dharmashankar Subramanian, Gaetano Rossiello,
  Long Vu, and Alfio Gliozzo. 2025.
\newblock \href {https://arxiv.org/abs/2501.17174} {{Extractive Schema Linking
  for Text-to-SQL}}.
\newblock \emph{Preprint}, arXiv:2501.17174.

\bibitem[{Hazoom et~al.(2021)Hazoom, Malik, and Bogin}]{hazoom2021sede}
Moshe Hazoom, Vibhor Malik, and Ben Bogin. 2021.
\newblock \href {https://aclanthology.org/2021.nlp4prog-1.9/} {{Text-to-SQL in
  the Wild: A Naturally-Occurring Dataset Based on Stack Exchange Data}}.
\newblock In \emph{Proceedings of the 1st Workshop on Natural Language
  Processing for Programming}.

\bibitem[{Jayaraman et~al.(2025)Jayaraman, Marathe, Mozaffari, Shen, and
  Kenthapadi}]{jayaraman2025permissioned}
Bargav Jayaraman, Virendra~J. Marathe, Hamid Mozaffari, William~F. Shen, and
  Krishnaram Kenthapadi. 2025.
\newblock \href {https://arxiv.org/abs/2505.22860} {{Permissioned LLMs:
  Enforcing Access Control in Large Language Models}}.
\newblock \emph{Preprint}, arXiv:2505.22860.

\bibitem[{Klisura et~al.(2025)Klisura, Khoury, Kundu, Krishnan, and
  Rios}]{klisura2025role}
Djordje Klisura, Joseph Khoury, Ashish Kundu, Ram Krishnan, and Anthony Rios.
  2025.
\newblock \href {https://arxiv.org/abs/2510.07642} {{Role-Conditioned Refusals:
  Evaluating Access Control Reasoning in Large Language Models}}.
\newblock \emph{Preprint}, arXiv:2510.07642.

\bibitem[{Agrawal and Li(2022)}]{agrawal2022debiasgan}
Adarsh Agrawal and Jessica Li. 2022.
\newblock {Mitigating Bias in AI Using Debias-GAN}.
\newblock White paper, World Wide Technology.

\bibitem[{Lei et~al.(2025)Lei, Chen, Ye, Cao, Shin, Su, Suo, Gao, Hu, Yin,
  Zhong, Xiong, Sun, Liu, Wang, and Yu}]{lei2025spider2}
Fangyu Lei, Jixuan Chen, Yuxiao Ye, Ruisheng Cao, Dongchan Shin, Hongjin Su,
  Zhaoqing Suo, Hongcheng Gao, Wenjing Hu, Pengcheng Yin, Victor Zhong, Caiming
  Xiong, Ruoxi Sun, Qian Liu, Sida Wang, and Tao Yu. 2025.
\newblock \href {https://mlanthology.org/iclr/2025/lei2025iclr-spider/}
  {{Spider 2.0: Evaluating Language Models on Real-World Enterprise Text-to-SQL
  Workflows}}.
\newblock In \emph{International Conference on Learning Representations}.

\bibitem[{Li et~al.(2023)Li, Hui, Qu, Yang, Li, Li, Wang, Qin, Geng, Huo
  et~al.}]{li2023bird}
Jinyang Li, Binyuan Hui, Ge~Qu, Jiaxi Yang, Binhua Li, Bowen Li, Bailin Wang,
  Bowen Qin, Ruiying Geng, Nan Huo, and 1 others. 2023.
\newblock \href
  {https://proceedings.neurips.cc/paper_files/paper/2023/hash/9ee883c8a46d6ac8747b4d6edc7e1a6b-Abstract-Datasets_and_Benchmarks.html}
  {{Can LLM Already Serve as A Database Interface? A BIg Bench for Large-Scale
  Database Grounded Text-to-SQL}}.
\newblock In \emph{Advances in Neural Information Processing Systems}.

\bibitem[{Liu et~al.(2025)Liu, Tan, Zhong, Xie, Zhao, Wang, Hu, and
  Li}]{liu2025solid}
Geling Liu, Yunzhi Tan, Ruichao Zhong, Yuanzhen Xie, Lingchen Zhao, Qian Wang,
  Bo~Hu, and Zang Li. 2025.
\newblock \href {https://aclanthology.org/2025.coling-main.654/} {{Solid-SQL:
  Enhanced Schema-linking based In-context Learning for Robust Text-to-SQL}}.
\newblock In \emph{Proceedings of the 31st International Conference on
  Computational Linguistics}.

\bibitem[{Luo et~al.(2023)Luo, Xie, Chen, He, Li, Chen, and
  Yang}]{luo2023crush4sql}
Zhiqiang Luo, Liang Xie, Jingping Chen, Yiduo He, Zhenyu Li, Weian Chen, and
  Bo~Yang. 2023.
\newblock \href {https://aclanthology.org/2023.emnlp-main.868/} {{CRUSH4SQL:
  Collective Retrieval Using Schema Hallucination For Text2SQL}}.
\newblock In \emph{Proceedings of the 2023 Conference on Empirical Methods in
  Natural Language Processing}.

\bibitem[{Nahid et~al.(2025)Nahid, Rafiei, Zhang, and
  Zhang}]{nahid2025rethinking}
Md~Mahadi~Hasan Nahid, Davood Rafiei, Weiwei Zhang, and Yong Zhang. 2025.
\newblock \href {https://arxiv.org/abs/2510.14296} {{Rethinking Schema Linking:
  A Context-Aware Bidirectional Retrieval Approach for Text-to-SQL}}.
\newblock \emph{Preprint}, arXiv:2510.14296.

\bibitem[{Pourreza and Rafiei(2023)}]{pourreza2023dinsql}
Mohammadreza Pourreza and Davood Rafiei. 2023.
\newblock \href {https://arxiv.org/abs/2304.11015} {{DIN-SQL: Decomposed
  In-Context Learning of Text-to-SQL with Self-Correction}}.
\newblock In \emph{Advances in Neural Information Processing Systems}.

\bibitem[{Safdarian et~al.(2025)Safdarian, Mohammadi, Jahanbakhsh,
  Shahamat~Naderi, and Faili}]{safdarian2025schemagraphsql}
AmirHossein Safdarian, Milad Mohammadi, Ehsan Jahanbakhsh, Mona
  Shahamat~Naderi, and Heshaam Faili. 2025.
\newblock \href {https://arxiv.org/abs/2505.18363} {{SchemaGraphSQL: Efficient
  Schema Linking with Pathfinding Graph Algorithms for Text-to-SQL on
  Large-Scale Databases}}.
\newblock \emph{Preprint}, arXiv:2505.18363.

\bibitem[{Sharma(2025)}]{sharma2025rag}
Chaitanya Sharma. 2025.
\newblock \href {https://arxiv.org/abs/2506.00054} {{Retrieval-Augmented
  Generation: A Comprehensive Survey of Architectures, Enhancements, and
  Robustness Frontiers}}.
\newblock \emph{Preprint}, arXiv:2506.00054.

\bibitem[{Upadhyay et~al.(2026)Upadhyay, Thakur, Pradeep, Craswell, Campos, and
  Lin}]{upadhyay2026trec}
Shivani Upadhyay, Nandan Thakur, Ronak Pradeep, Nick Craswell, Daniel Campos,
  and Jimmy Lin. 2026.
\newblock \href {https://arxiv.org/abs/2603.09891} {{Overview of the TREC 2025
  Retrieval Augmented Generation Track}}.
\newblock \emph{Preprint}, arXiv:2603.09891.

\bibitem[{Vidgen et~al.(2024)Vidgen, Agrawal, Ahmed, Akinwande, Al-Nuaimi,
  Alfaraj, Alhajjar, Aroyo, Bavalatti, Blili-Hamelin
  et~al.}]{vidgen2024mlcommons}
Bertie Vidgen, Adarsh Agrawal, Ahmed~M. Ahmed, Victor Akinwande, Namir
  Al-Nuaimi, Najla Alfaraj, Elie Alhajjar, Lora Aroyo, Trupti Bavalatti,
  Borhane Blili-Hamelin, and 1 others. 2024.
\newblock \href {https://arxiv.org/abs/2404.12241} {{Introducing v0.5 of the AI
  Safety Benchmark from MLCommons}}.
\newblock \emph{arXiv preprint arXiv:2404.12241}.

\bibitem[{Wang et~al.(2025{\natexlab{a}})Wang, Ren, Yang, Liang, Bai, Zhang,
  Yan, and Li}]{wang2024macsql}
Bing Wang, Changyu Ren, Jian Yang, Xinnian Liang, Jiaqi Bai, Linzheng Zhang,
  Zhao Yan, and Zhoujun Li. 2025{\natexlab{a}}.
\newblock \href {https://arxiv.org/abs/2312.11242} {{MAC-SQL: A Multi-Agent
  Collaborative Framework for Text-to-SQL}}.
\newblock In \emph{Proceedings of the 31st International Conference on
  Computational Linguistics}.

\bibitem[{Wang et~al.(2025{\natexlab{b}})Wang, Liu, and
  Yang}]{wang2025linkalign}
Yihan Wang, Peiyu Liu, and Xin Yang. 2025{\natexlab{b}}.
\newblock \href {https://aclanthology.org/2025.emnlp-main.51/} {{LinkAlign:
  Scalable Schema Linking for Real-World Large-Scale Multi-Database
  Text-to-SQL}}.
\newblock In \emph{Proceedings of the 2025 Conference on Empirical Methods in
  Natural Language Processing}.

\bibitem[{Wang et~al.(2026)Wang, Zheng, Cao, Zhang, Wei, Fu, Luo, Chen, and
  Bai}]{wang2026autolink}
Ziyang Wang, Yuanlei Zheng, Zhenbiao Cao, Xiaojin Zhang, Zhongyu Wei, Pei Fu,
  Zhenbo Luo, Wei Chen, and Xiang Bai. 2026.
\newblock \href {https://ojs.aaai.org/index.php/AAAI/article/view/40672}
  {{AutoLink: Autonomous Schema Exploration and Expansion for Scalable Schema
  Linking in Text-to-SQL at Scale}}.
\newblock In \emph{Proceedings of the AAAI Conference on Artificial
  Intelligence}.

\end{thebibliography}

\clearpage
\appendix
\begin{center}
{\Large\bfseries Appendix}
\end{center}
\vspace{0.3em}

This appendix provides full retrieval sweeps (\secref{app:full_results}), implementation details and hyperparameters (\secref{app:implementation}), the exact text templates used to render catalog objects for embedding (\secref{app:templates}), additional analyses including query complexity, BIRD difficulty stratification, signal weight sensitivity, retrieval-architecture ablations, retrieval failure analysis, and catalog metadata noise robustness (\secref{app:additional}, \secref{app:noise}), and artifact notes for reproducibility (\secref{app:artifact_notes}).

\vspace{0.5em}
\FloatBarrier

\section{Full Retrieval Results}
\label{app:full_results}

This section reports complete retrieval sweeps across all evaluated $K$ values for both CRUSH4SQL and SEDE. The summary in \tabref{tab:retrieval_main} of the main paper reports a small slice of these results; the full sweeps make explicit which improvements affect early precision and which affect high-coverage retrieval. We examine each benchmark in turn.

\textbf{CRUSH4SQL.}

\begin{table}[t]
\centering
\small
\setlength{\tabcolsep}{2.2pt}
\begin{tabular}{@{}lcccccc@{}}
\toprule
\textbf{Metric} & \textbf{1} & \textbf{3} & \textbf{5} & \textbf{10} & \textbf{20} & \textbf{50} \\
\midrule
Table, no & 51.2 & 72.4 & 82.8 & 89.7 & 94.3 & 97.1 \\
Table, +R & 62.8 & 81.3 & 89.1 & 93.5 & 96.4 & 98.2 \\
\midrule
Column, no & 14.2 & 24.8 & 33.4 & 44.4 & 55.2 & 68.9 \\
Column, +R & 21.7 & 34.5 & 43.9 & 55.5 & 64.1 & 76.3 \\
\bottomrule
\end{tabular}
\caption{\textbf{Full CRUSH4SQL recall sweeps} (1,534 questions). Columns are Recall@$K$.}
\label{tab:crush_full}
\end{table}

At the table level, the reranker improves recall at $K=1$ by +11.6 absolute points (51.2\% to 62.8\%), indicating that the cross-encoder frequently promotes the correct table from lower bi-encoder positions to the top. At $K=50$, both conditions approach ceiling performance (97.1\% vs.\ 98.2\%), confirming that the bi-encoder provides sufficient coverage at high $K$ and the reranker's primary contribution is precision at low $K$. Column retrieval exhibits substantially lower absolute performance than table retrieval across all conditions, reflecting the inherent difficulty of matching natural language to abbreviated column names.

\textbf{SEDE.}

\begin{table}[t]
\centering
\small
\setlength{\tabcolsep}{2.2pt}
\begin{tabular}{@{}lcccccc@{}}
\toprule
\textbf{Metric} & \textbf{1} & \textbf{3} & \textbf{5} & \textbf{10} & \textbf{20} & \textbf{50} \\
\midrule
Table, no & 22.3 & 41.8 & 52.1 & 71.2 & 83.5 & 91.4 \\
Table, +H & 74.1 & 87.6 & 92.3 & 94.2 & 95.0 & 96.8 \\
\midrule
Single, no & 38.5 & 59.2 & 71.0 & 82.4 & 89.7 & 94.3 \\
Single, +H & 89.2 & 95.7 & 98.4 & 99.1 & 99.5 & 99.8 \\
\bottomrule
\end{tabular}
\caption{\textbf{Full SEDE table recall sweeps} (857 questions). Columns are Recall@$K$.}
\label{tab:sede_full}
\end{table}

At $K=50$, the gap between +History and No History narrows to 5.4 points (96.8\% vs.\ 91.4\%), confirming that the bi-encoder eventually retrieves the correct tables given sufficient budget. History's contribution is therefore primarily one of ranking precision: it achieves at $K=5$ what the bi-encoder alone requires $K=20$ to approach. For the subset of queries requiring exactly one table (approximately 62\% of SEDE), history achieves 98.4\% recall at $K=5$, showing that query history strongly covers direct table retrieval.

\section{Implementation Details}
\label{app:implementation}

This section provides reproducible implementation details. All hyperparameters were selected on held-out development splits and frozen before final evaluation. The full hyperparameter table appears at the end of this section (\tabref{tab:hyperparams}). We describe the embedding infrastructure, reranker configuration, and SQL-generation setup in turn.

\textbf{Embedding infrastructure.} The implementation uses Amazon Titan Embeddings V2 via AWS Bedrock. The model produces 1,024-dimensional vectors with a maximum input length of 8,192 tokens. All embeddings are normalized to unit length for cosine similarity. We use the default model configuration without task-specific fine-tuning. For local vector storage we use ChromaDB with HNSW indexing. The HNSW parameters are ef\_construction=200, M=32, and ef\_search=100. The vector indexes are organized by catalog object type. At query time, all five indexes are searched in parallel and results are merged. This separation enables type-specific retrieval thresholds and simplifies access-control filtering because each index can be independently filtered.

\textbf{Reranking.} The reranker is Cohere Rerank V3.5 through AWS Bedrock. Each catalog object is rendered to its text template and truncated to 512 tokens before reranking. Candidates from all object types are pooled and reranked jointly.

\textbf{SQL generation.} SQL generation uses the repository default Claude Haiku-class model for simple questions and the same prompt template across compared conditions. Generation temperature is set to 0, with maximum output length 2,048 tokens.

\begin{table}[t]
\centering
\small
\begin{tabular}{@{}lll@{}}
\toprule
\textbf{Component} & \textbf{Parameter} & \textbf{Value} \\
\midrule
Bi-encoder & Top per type & 100 \\
Bi-encoder & Min similarity & 0.3 \\
Bi-encoder & Table limit & 50 \\
Bi-encoder & Column limit & 100 \\
\midrule
Reranker & Pool size & 100 \\
Reranker & Max doc length & 512 tokens \\
\midrule
Fusion & Embedding weight $\alpha$ & 0.3 \\
Fusion & Reranker weight $\beta$ & 0.5 \\
Fusion & History weight $\gamma$ & 0.2 \\
\midrule
History & Decay half-life & 30 days \\
History & Max patterns per DB & 1,000 \\
History & Min execution count & 2 \\
\midrule
Output & Final tables & Top 20 \\
Output & Final columns & Top 50 \\
\midrule
SQL generation & Temperature & 0 \\
SQL gen. & Max output & 2,048 \\
\bottomrule
\end{tabular}
\caption{\textbf{Hyperparameter specification.} Values were selected on held-out development splits using grid search for fusion weights and threshold tuning for retrieval.}
\label{tab:hyperparams}
\end{table}

\section{Catalog Object Templates}
\label{app:templates}

This section gives the text templates used for the five catalog object types. The templates are rendered before embedding. They place discriminative fields early, use labeled natural-language fields rather than raw key-value dumps, and keep rendered text under the reranker document length.

\subsection{Design Principles}

Template design follows three rules. First, discriminative fields (name, description) are placed at the top so that truncation during embedding preserves the most important retrieval signal. Second, templates use labeled natural-language fields rather than raw key-value dumps, aligning with the embedding model's training distribution. Third, all rendered templates stay under 512 tokens so that no truncation occurs during embedding or reranking.

\subsection{Table Template}

\begin{table}[t]
\centering
\scriptsize
\begin{tabular}{@{}l@{}}
\toprule
\ttfamily Table: \{\{ schema \}\}.\{\{ table\_name \}\} \\
\ttfamily Description: \{\{ description \}\} \\
\ttfamily Columns: \{\{ column\_count \}\} columns \\
\ttfamily Rows: \{\{ row\_count $|$ format\_number \}\} rows \\
\ttfamily Tags: \{\{ tags $|$ join(', ') \}\} \\
\ttfamily Owner: \{\{ owner\_team \}\} \\
\ttfamily Update Frequency: \{\{ update\_frequency \}\} \\
\bottomrule
\end{tabular}
\end{table}

The description and tags are placed early because they carry the strongest semantic content; row counts and ownership appear at the end as auxiliary disambiguation.

\subsection{Column Template}

\begin{table}[t]
\centering
\scriptsize
\begin{tabular}{@{}l@{}}
\toprule
\ttfamily Column: \{\{ table\_name \}\}.\{\{ column\_name \}\} \\
\ttfamily Type: \{\{ data\_type \}\} \\
\ttfamily Description: \{\{ description \}\} \\
\ttfamily Sample Values: \{\{ sample\_values[:5] $|$ join(', ') \}\} \\
\ttfamily Null Percentage: \{\{ null\_pct \}\}\% \\
\ttfamily Distinct: \{\{ distinct\_count $|$ format\_number \}\} \\
\ttfamily Is Primary Key: \{\{ is\_pk \}\} \\
\ttfamily Is Foreign Key: \{\{ is\_fk \}\} \\
\bottomrule
\end{tabular}
\end{table}

Sample values are the most discriminative field for entity-grounded queries (e.g., ``sales in California'' matches a state column with ``California'' in samples). Distinct count distinguishes identifier columns from categorical ones.

\subsection{Metric Template}

\begin{table}[t]
\centering
\scriptsize
\begin{tabular}{@{}l@{}}
\toprule
\ttfamily Metric: \{\{ metric\_name \}\} \\
\ttfamily Business Definition: \{\{ definition \}\} \\
\ttfamily SQL Formula: \{\{ sql\_formula \}\} \\
\ttfamily Sources: \{\{ source\_tables $|$ join(', ') \}\} \\
\ttfamily Columns: \{\{ source\_columns $|$ join(', ') \}\} \\
\ttfamily Unit: \{\{ unit \}\} \\
\ttfamily Aggregation: \{\{ aggregation\_type \}\} \\
\ttfamily Filters: \{\{ default\_filters $|$ join('; ') \}\} \\
\bottomrule
\end{tabular}
\end{table}

Metrics serve as bidirectional bridges: retrievable by business term (definition) and by technical reference (SQL formula). Including both expands the entry surface for the embedding.

\subsection{Relationship Template}

\begin{table}[t]
\centering
\scriptsize
\begin{tabular}{@{}l@{}}
\toprule
\ttfamily Relationship: \\
\ttfamily \quad \{\{ table\_a \}\}.\{\{ col\_a \}\} $\to$ \{\{ table\_b \}\}.\{\{ col\_b \}\} \\
\ttfamily Type: \{\{ relationship\_type \}\} \\
\ttfamily Cardinality: \{\{ cardinality \}\} \\
\ttfamily Description: \{\{ description \}\} \\
\ttfamily Join Condition: \{\{ join\_sql \}\} \\
\ttfamily Is Validated: \{\{ is\_validated \}\} \\
\ttfamily Usage Count: \{\{ usage\_count \}\} \\
\bottomrule
\end{tabular}
\end{table}

Usage count enables prioritization of frequently-used joins; the validation flag distinguishes verified foreign keys from documented but unvalidated relationships.

\subsection{Query History Template}

\begin{table}[t]
\centering
\scriptsize
\begin{tabular}{@{}l@{}}
\toprule
\ttfamily Historical Query Pattern: \\
\ttfamily Question: \{\{ natural\_language\_question \}\} \\
\ttfamily Tables Accessed: \{\{ tables $|$ join(', ') \}\} \\
\ttfamily Columns Used: \{\{ columns $|$ join(', ') \}\} \\
\ttfamily Join Pattern: \{\{ join\_pattern \}\} \\
\ttfamily Aggregations: \{\{ aggregations $|$ join(', ') \}\} \\
\ttfamily Filters: \{\{ filter\_columns $|$ join(', ') \}\} \\
\ttfamily Frequency: \{\{ execution\_count \}\} times \\
\ttfamily Last Used: \{\{ last\_executed \}\} \\
\bottomrule
\end{tabular}
\end{table}

Placing the natural-language question first enables direct semantic match against new queries. Frequency and last-used metadata feed the exponential decay weighting.

\section{Additional Analyses}
\label{app:additional}

This section presents six ablations and analyses that complement the main results: query complexity (\tabref{tab:complexity}), BIRD difficulty stratification (\tabref{tab:bird_difficulty}), signal weight sensitivity (\tabref{tab:weights}), retrieval-architecture ablation including BM25 and a unified-index variant (\tabref{tab:retrieval_ablation}), retrieval failure analysis on the per-question reranked output, and catalog metadata noise robustness (\figref{fig:noise}). Together they show that the system degrades gracefully on harder queries, generalizes across difficulty bands, is insensitive to exact weight choices, beats sparse retrieval by a large margin, gains from typed-separated indexes, and is robust to substantial drops in catalog metadata richness.

\subsection{Query complexity}

\begin{table}[t]
\centering
\small
\begin{tabular}{lcc}
\toprule
\textbf{Gold tables} & \textbf{Table R@5} & \textbf{Column R@10} \\
\midrule
1 table & 96.9\% & 73.4\% \\
2 tables & 88.5\% & 51.3\% \\
3 to 4 tables & 79.7\% & 45.0\% \\
\bottomrule
\end{tabular}
\caption{\textbf{Retrieval quality by query complexity} on CRUSH4SQL with reranking. Column recall falls more sharply than table recall as the query spreads across multiple tables.}
\label{tab:complexity}
\end{table}

\begin{figure}[t]
\centering
\includegraphics[width=0.7\columnwidth]{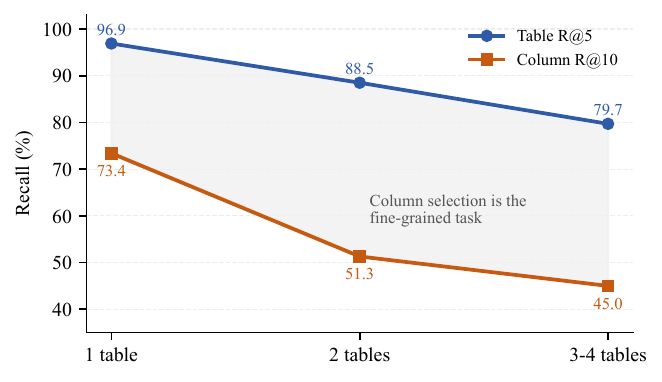}
\caption{\textbf{Retrieval performance by query complexity} (number of gold tables required). Single-table queries achieve near-perfect table retrieval (96.9\% R@5). Column recall degrades more steeply under multi-table queries, identifying fine-grained column retrieval as the primary remaining challenge.}
\label{fig:complexity}
\end{figure}

\textbf{Single-table queries are nearly solved.} With one gold table, the system reaches 96.9\% table recall@5 and 73.4\% column recall@10. As the number of required tables grows, table recall declines gracefully ($-$17.2 points from 1 to 3-4 tables), but column recall falls more sharply ($-$28.4 points). The reason is structural: a single-table query needs columns from one table, while a 4-table query must distribute the same column-retrieval budget across four tables. This is why relationship objects and lineage expansion are part of the retrieval path rather than a stylistic addition; they bring join-key evidence into context exactly when single-table column retrieval is no longer sufficient.

\subsection{BIRD difficulty stratification}

\begin{table}[t]
\centering
\small
\begin{tabular}{@{}lcccc@{}}
\toprule
\textbf{Diff.} & \textbf{Full} & \textbf{SFR} & \textbf{$\Delta$} & \textbf{N} \\
\midrule
Simple & 89.7 & 93.1 & +3.4 & 58 \\
Moderate & 75.0 & 96.9 & +21.9 & 32 \\
Chall. & 83.3 & 83.3 & +0.0 & 6 \\
\midrule
Overall & 84.4 & 93.8 & +9.4 & 96 \\
\bottomrule
\end{tabular}
\caption{\textbf{BIRD execution accuracy by difficulty level.} Both conditions use the same model and prompt. The largest gain appears on moderate queries.}
\label{tab:bird_difficulty}
\end{table}

\textbf{Moderate queries gain most.} The largest execution-accuracy gain (+21.9 points) appears on moderate queries, where schema distractors are common enough to hurt full-schema prompting but the required joins and filters remain recoverable once context is focused. Simple queries already gain less (+3.4) because the correct table is often obvious even with distractors. The challenging split contains only six questions in this 96-question subset, so its flat result should be read as neutral rather than negative.

\subsection{Signal weight sensitivity}

\begin{table}[t]
\centering
\footnotesize
\setlength{\tabcolsep}{4pt}
\begin{tabular}{ccccc}
\toprule
$\alpha$ & $\beta$ & $\gamma$ & \textbf{Table R@5} & \textbf{Col R@10} \\
\midrule
1.0 & 0.0 & 0.0 & 82.8\% & 44.4\% \\
0.0 & 1.0 & 0.0 & 87.3\% & 53.8\% \\
0.5 & 0.5 & 0.0 & 88.6\% & 54.9\% \\
0.3 & 0.5 & 0.2 & 89.1\% & 55.5\% \\
0.2 & 0.6 & 0.2 & 88.9\% & 55.1\% \\
0.3 & 0.3 & 0.4 & 88.4\% & 53.2\% \\
\bottomrule
\end{tabular}
\caption{\textbf{Signal weight sensitivity on the CRUSH4SQL development set.} The selected configuration assigns highest weight to reranker relevance, moderate weight to embedding similarity, and lower weight to query history.}
\label{tab:weights}

\vspace{0.6em}

\centering
\footnotesize
\setlength{\tabcolsep}{2pt}
\begin{tabular}{@{}>{\raggedright\arraybackslash}p{0.52\columnwidth}cc@{}}
\toprule
\textbf{Condition} & \textbf{Table R@5} & \textbf{Col R@10} \\
\midrule
BM25 over rendered templates & 50.0\% & 28.7\% \\
Unified single index (5 types merged) & 78.3\% & 45.0\% \\
Typed-separated indexes (ours) & 80.7\% & 44.9\% \\
\bottomrule
\end{tabular}
\caption{\textbf{Retrieval-architecture ablation on CRUSH4SQL.} BM25 over the same rendered templates loses 30 points at Table R@5 to the bi-encoder, confirming that semantic similarity beyond lexical overlap is what the embedding pipeline contributes. Merging all five typed catalog collections into a single index with a type-tag prefix loses 2.4 points at Table R@5; typed separation lets the bi-encoder concentrate on per-type semantics during the initial broad search.}
\label{tab:retrieval_ablation}
\end{table}

\textbf{The system is robust to weight tuning.} The reranker alone ($\alpha{=}0,\beta{=}1,\gamma{=}0$) outperforms the bi-encoder alone by +4.5 table R@5 and +9.4 column R@10, confirming that cross-encoder attention provides stronger relevance signal than independent embedding comparison. The combined system gains another +1.8 / +1.7 points over the best single signal. The top four configurations span only 0.7 points in table R@5, indicating that exact weight tuning is not required: approximate weights selected on a small development set transfer effectively to the full evaluation.

\subsection{Retrieval-architecture ablation}

\tabref{tab:retrieval_ablation} reports two architectural ablations against the typed-separated pipeline. \textbf{BM25 baseline.} Sparse retrieval over the same rendered catalog templates reaches only 50.0\% Table R@5 and 28.7\% Column R@10, $-$30.7 and $-$16.2 points relative to the typed-separated pipeline. The gap quantifies the contribution of semantic embeddings over lexical overlap and confirms that the gains are not just from the templates. \textbf{Unified single index.} Merging the five typed collections into a single index with a `[TYPE=table]'-style tag prefix and using the same Cohere reranker yields $-$2.4 points at Table R@5 and $-0.1$ points at Column R@10. The column-level result is essentially flat because the cross-encoder reranker, which we apply jointly across types in both conditions, dominates fine-grained column ordering. The Table R@5 gap shows that typed separation lets the bi-encoder concentrate on per-type semantics during the broad initial search, before the reranker is invoked. Both ablations were run on the same 1{,}534-question CRUSH4SQL split with a heuristic intent classifier (no LLM) for cost-controlled comparison; absolute numbers are therefore lower than the production pipeline's 89.1\% Table R@5 in \tabref{tab:retrieval_main}, but deltas are directly comparable within this set.

\subsection{Retrieval failure analysis}
\label{app:failure_analysis}

To understand where retrieval fails on CRUSH4SQL, we categorize per-gold-table misses on the reranked baseline (89.1\% Table R@5). Of 2{,}969 (question, gold-table) pairs across 1{,}534 questions, 377 (12.7\%) miss the top-5 ranking. Failures stratify cleanly by the number of gold tables required: 3.1\% fail rate for single-table queries, 22.1\% for 2-table, 48.9\% for 3-table, 68.2\% for 4-table. The combinatorial budget pressure of fitting all gold tables into a fixed-size top-5 explains most of the multi-table degradation.

We classify each miss into multi-label categories. The dominant failure modes are: (i) abbreviation/lexical mismatch (323 of 377 misses, 86\%) where the gold table name has low overlap with the question tokens and the description is not informative enough; (ii) ambiguous topic (224 of 377, 59\%) where multiple plausible tables in the gold table's database appear above the gold table; (iii) cross-domain confusion (23 misses, 6\%) where the top-5 contains zero tables from the gold's database, typically when a question phrased generically (``how many users'') matches similar tables in unrelated databases; (iv) sparse description (14 misses, 4\%); (v) multi-table dispersion (21 misses, where 4+ gold tables are required). The two dominant categories point at the same underlying lever: richer per-table signal (descriptions, tags, sample values) would reduce both abbreviation mismatch and ambiguous-topic confusion. The dependency on metadata richness is why we ran the noise robustness study in \secref{app:noise}.

\subsection{Catalog metadata noise robustness}
\label{app:noise}

A reasonable concern with catalog-first retrieval is that production catalogs are sparser than benchmark catalogs. We probe this directly: starting from the typed-separated pipeline, we randomly drop 25\%, 50\%, and 75\% of column descriptions and sample values (and table descriptions) and re-run the full CRUSH4SQL evaluation.

\begin{figure}[t]
\centering
\includegraphics[width=0.95\columnwidth]{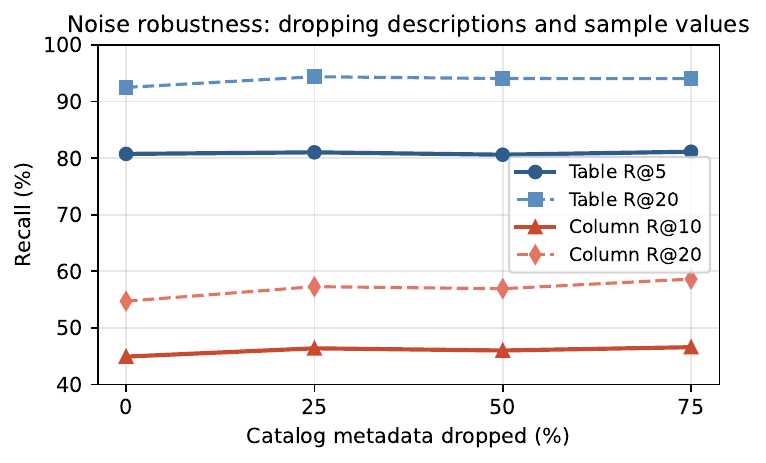}
\caption{\textbf{Noise robustness on CRUSH4SQL.} Recall stays within $\pm$2 points across 0/25/50/75\% metadata drop. The system is essentially noise-immune on this benchmark because CRUSH4SQL table and column NAMES carry most of the discriminative signal; the descriptions in CRUSH4SQL's union schema are minimally populated to begin with.}
\label{fig:noise}
\end{figure}

\textbf{Result.} \figref{fig:noise} shows that Table R@5 stays in $[80.6, 81.1]$ and Column R@10 stays in $[44.9, 46.6]$ across all four conditions, including 75\% metadata drop. The lack of degradation is benchmark-specific: CRUSH4SQL's union schema populates only minimal descriptions to begin with, so the bi-encoder's reliance on table and column NAMES makes it robust to additional description loss. On a deployment catalog with rich, hand-written descriptions, we would expect noise sensitivity to be higher; we flag this as a limitation in the main text. The result does, however, support the more limited claim that the system does not collapse when descriptions are imperfect, which is a useful deployment guarantee.

\section{Artifact Notes}
\label{app:artifact_notes}

\textbf{Components.} The artifact is organized around typed catalog models, pluggable embedding providers, vector-store backends (a development backend for local evaluation and a production backend for scaled deployments), the retrieval orchestrator, the lineage graph, the SQL generator, the post-generation verifier, and the deterministic access-control rewriter described in \secref{subsec:generation_acl}.

\textbf{Evaluation harness.} The harness includes adapters for CRUSH4SQL, SEDE, and BIRD that convert each benchmark into a uniform catalog representation and an evaluation question set. Retrieval metrics (Recall@K, MRR, NDCG@10, MAP@10) are computed from the harness output; SQL-generation accuracy is computed by executing both predicted and gold SQL against the benchmark databases and comparing result sets.

\textbf{Intended use.} The artifact is intended for research evaluation of catalog-first retrieval for natural language analytics. Deployment on private catalogs should apply the authorization, audit, retention, and minimization guidance discussed in the Ethical Considerations section of the main paper.

\end{document}